# Counting energy packets in the electromagnetic wave

## Stefan Popescu[1] and Bernhard Rothenstein[2]

1) Siemens AG, Erlangen, Germany stefan.popescu@siemens.com
2) Politehnica University of Timisoara, Physics Department,
Timisoara, Romania brothenstein@gmail.com

*Abstract. We discuss the concept of energy packets in respect to the energy transported by electromagnetic waves and we demonstrate that this physical quantity can be used in physical problems involving relativistic effects. This refined concept provides results compatible to those obtained by simpler definition of energy density when relativistic effects apply to the free electromagnetic waves. We found this concept further compatible to quantum theory perceptions and we show how it could be used to conciliate between different physical approaches including the classical electromagnetic wave theory, the special relativity and the quantum theories.*

## 1. Introduction to electromagnetic wave theory

The plane electromagnetic wave travels in vacuum as fleeting orthogonal electric and magnetic fields. The field intensities oscillate spatially along the direction of propagation and temporary normal to the direction of propagation as depicted by figure 1.

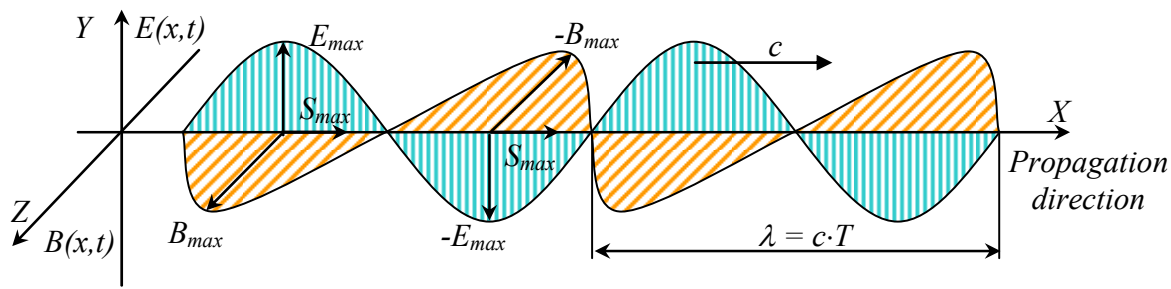

*Figure 1. Time-space propagation of the plane electromagnetic wave in vacuum depicts the orthogonal electric and magnetic components that oscillate both temporary at any point with period T and longitudinally at any time instant with period λ.*

A stationary observer located at some arbitrary point *x* along the propagation direction measures the local amplitude of the electric and magnetic field and records synchronous electric and magnetic oscillations having the same period *T* and the same phase. A group of observers equipped with synchronized watches and distributed along the propagation direction measure the local amplitude of the electric and magnetic field. When comparing their measurements made at any arbitrary time *t* these observers identify a longitudinally oscillating electric and magnetic field having the same period $\lambda = c \cdot T$ and the same phase. In vacuum these fleeting oscillations are mathematically described by the familiar wave equations:

$$E(x,t) = E_{max} \sin \omega \left(t - \frac{x}{c}\right) \quad B(x,t) = B_{max} \sin \omega \left(t - \frac{x}{c}\right) \tag{1}$$

and we have further $E(x,t) = cB(x,t)$ with $c = \frac{1}{\sqrt{\varepsilon_0 \mu_0}}$, $B(x,t) = \mu_0 H(x,t)$ and

$Z_{m0} = \frac{E(x,t)}{H(x,t)} = \sqrt{\frac{\mu_0}{\varepsilon_0}} \cong 377.79\,\Omega$ being the intrinsic vacuum impedance.

The energy flow with the electromagnetic wave
The directional energy flow in the wave is further given by the Poynting vector expressed as the wave power deposed per unit area:

$$\vec{S}(x,t) = \varepsilon_0 c^2 \vec{E}(x,t) \times \vec{B}(x,t) \tag{2}$$

In a plane wave this vector is always pointing in the direction of wave propagation and in vacuum its amplitude is given by the expression:

$$S(x,t) = c \cdot \rho(x,t) = c\varepsilon_0 E^2(x,t) \tag{3}$$

Figure 1 shows the Poynting vector $S$ maintaining the direction but changing periodically in amplitude from peak values at wave crests to zero values in between.

The energy density in the wave space
The punctual and instantaneous energy density in the wave space is given by the well known equation:

$$\rho(x,t) = \frac{1}{2}\left[\varepsilon_0 E^2(x,t) + \mu_0 H^2(x,t)\right] = \frac{1}{2}\varepsilon_0 \left[E^2(x,t) + c^2 B^2(x,t)\right] \tag{4}$$

In vacuum this relation reduces to simple forms:

$$\rho(x,t) = \varepsilon_0 E^2(x,t) = \mu_0 H^2(x,t) = \varepsilon_0 c^2 B^2(x,t) \tag{5}$$

Replacing the expressions for wave fields (1) we get the explicit form as:

$$\rho(x,t) = \varepsilon_0 E_{max}^2 \sin^2 \omega \left(t - \frac{x}{c}\right) = \rho_{max} \sin^2 \omega \left(t - \frac{x}{c}\right) \tag{6}$$

with $\rho_{max} = \varepsilon_0 E_{max}^2$. This equation shows that the energy carried by the plane wave oscillates temporary at any arbitrary location and longitudinally at any arbitrary time instant. The energy density changes between zero and a peak value also called wave crest as further symbolically depicted by figure 2. This figure represents intuitively the spatial-temporal oscillations of the wave energy density as a travelling tread of spindles.

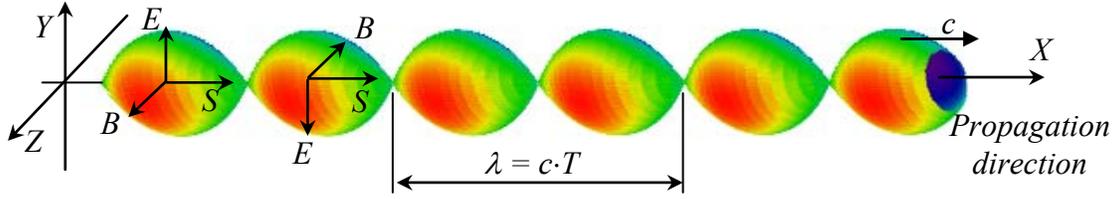

***Figure 2***. *Time-Space propagation of the plane electromagnetic wave depicts energy packets repeating spatially at half the wavelength period λ/2 and temporary at half the wave period T/2*

The mean energy density in any arbitrary point $x$ along the propagation axis is given by a temporal integration of (4) as:

$$\rho_{med}(x) = \frac{2}{T} \int_{-T/4}^{T/4} \rho_{max} \sin^2 \omega\left(t - \frac{x}{c}\right) dt = \frac{\rho_{max}}{2} = \frac{\varepsilon_0 E_{max}^2}{2} \qquad (7)$$

In a non-dissipative propagation medium the mean energy density is the same in any point. We calculate also the mean energy density in the volume between two wave crests at any arbitrary time instant as:

$$\rho_{med}(t) = \frac{2}{\lambda} \int_{-\lambda/4}^{\lambda/4} \rho_{max} \sin^2 \omega\left(t - \frac{x}{c}\right) dt = \frac{\rho_{max}}{2} = \frac{\varepsilon_0 E_{max}^2}{2} \qquad (8)$$

In a non-dissipative propagation medium the mean energy density is time invariant.

## 2. The concept of wave energy packet in the classic approach

The synchronous oscillations of the electric and magnetic fields result in periodical oscillations of wave energy density. An observer using an electric antenna and a wire loop to probe these fields detects short time windows when either probe hardly outputs some signal (insignificant wave energy is detected) alternating with time windows showing maximum signals (peak wave energy is detected). By analogy with ocean waves the energy peaks are called wave crests and the wave energy travels as a succession of wave crests. This behaviour supports us to assume that that the wave energy is travelling as a succession of discrete energy pulses that we call wave energy packets - *WEP* and we associate a *WEP* to every wave crest. At any moment a *WEP* is delimited longitudinally by the two points flanking a wave crest where the energy density reaches zero. In each point along the propagation direction the *WEP* is delimited temporary by the two time instants flanking the crest arrival time when the energy density just passes through zero. Therefore the *WEP* is half wavelength long and the *WEP* transit time is half the wave period. The concept of wave energy packet is used here to delimit temporary or spatially the periodic variations in energy density and should not be yet confronted with the wave packet concept as known from quantum mechanics.

We are now concerned with calculating the mean energy that the plane wave transports by means of a single *WEP* throughout a stationary surface of area *A* orthogonal to the direction of wave propagation at some arbitrary position *x*. This may be for example the mean energy stored into a photo-detector having a transversal sensitive window of area *A* subsequent to the absorption of all light photons within a single wave packet. We calculate the mean energy by a time integrating of (4) for the duration of the *WEP*. In a non-dissipative propagation medium we have no attenuation during propagation and therefore the mean energy in the *WEP* is the same in any point along the propagation direction.

$$W_{packet} = \int_0^{T/2} \rho(x,t) Ac \, dt = Ac\varepsilon_0 \int_0^{T/2} E^2(x,t) dt = AcT\varepsilon_0 \frac{E_{max}^2}{4} = A\lambda\varepsilon_0 \frac{E_{max}^2}{4} \qquad (9)$$

For consistency we obtain the same result by a spatial integration along the packet length at any arbitrary time instant *t*:

$$W_{packet} = \int_0^{\lambda/2} \rho(x,t) A \, dx = A\varepsilon_0 \int_0^{\lambda/2} E^2(x,t) dx = A\lambda\varepsilon_0 \frac{E_{max}^2}{4} = AcT\varepsilon_0 \frac{E_{max}^2}{4} \qquad (10)$$

## 3. The electromagnetic wave energy packet in the relativistic approach

In this approach we are concerned with the wave energy packets and how they are seen from two inertial reference frames K and K' in the standard arrangement as depicted in figure 3. A light source *S* stationary in reference frame K emits a plane wave consisting of successive energy packets that are measured by a photo-detector PD stationary in frame K and also by a photo-detector PD' moving in respect with *S* in the positive direction of OX axis.

The Lorentz transformations for the electromagnetic field components are:

$$E'_x = E_x \qquad E'_y = \frac{E_y - V \cdot B_z}{\sqrt{1-\beta^2}} \qquad B'_y = \frac{B_y + \frac{V}{c^2} \cdot E_z}{\sqrt{1-\beta^2}} \qquad (11)$$

$$B'_x = B_x \qquad E'_z = \frac{E_z + V \cdot B_y}{\sqrt{1-\beta^2}} \qquad B'_z = \frac{B_z - \frac{V}{c^2} \cdot E_y}{\sqrt{1-\beta^2}}$$

where $\beta = V/c$. For the plane and linear polarised wave we have $E_x = B_x = E_z = B_y = 0$ and therefore $E'_x = B'_x = E'_z = B'_y = 0$ thus when seen from K' the wave remains also linear polarised with the wave field intensities given by:

$$E'(x',t') = \frac{E(x,t) - V \cdot B(x,t)}{\sqrt{1-\beta^2}} \qquad B'(x',t') = \frac{B(x,t) - \frac{V}{c^2} \cdot E(x,t)}{\sqrt{1-\beta^2}} \qquad (12)$$

Using the expressions (1) for the stationary field components under the vacuum condition $E(x,t) = cB(x,t)$ and the inverse Lorenz transformation for the space and time coordinates

$$x = \frac{1}{\sqrt{1-\beta^2}}(x' + Vt') \qquad t = \frac{1}{\sqrt{1-\beta^2}}(t' + \frac{V}{c^2} \cdot x') \qquad (13)$$

we obtain the expression of wave components as seen by observers in K' as:

$$E'(x',t') = \sqrt{\frac{1-\beta}{1+\beta}} E_{max} \sin \omega \sqrt{\frac{1-\beta}{1+\beta}}\left(t' - \frac{x'}{c}\right)$$

$$B'(x',t') = \sqrt{\frac{1-\beta}{1+\beta}} B_{max} \sin \omega \sqrt{\frac{1-\beta}{1+\beta}}\left(t' - \frac{x'}{c}\right) \qquad (14)$$

With $E'_{max} = \sqrt{\frac{1-\beta}{1+\beta}} E_{max}$, $B'_{max} = \sqrt{\frac{1-\beta}{1+\beta}} B_{max}$ and $\omega' = \omega\sqrt{\frac{1-\beta}{1+\beta}}$ we obtain:

$$E'(x',t') = E'_{max} \sin \omega'\left(t' - \frac{x'}{c}\right) \qquad B'(x',t') = B'_{max} \sin \omega'\left(t' - \frac{x'}{c}\right) \qquad (15)$$

Therefore for the moving observers the wave remains plain polarised but the wave frequency and the amplitude of the electric and magnetic oscillations decrease by the Doppler factor. Correspondingly the wavelength increases by the Doppler factor and the energy density decreases by the square of the Doppler factor.

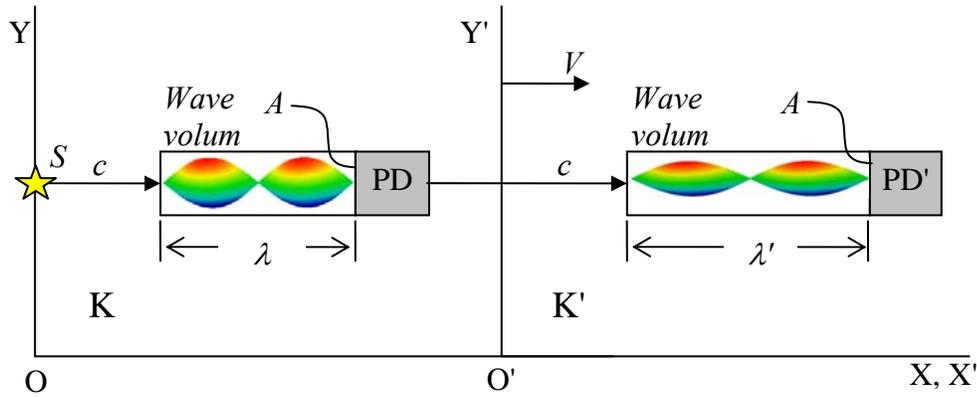

*Figure 3. The wave energy packets as measured by two similar photo detectors - each stationary in its reference frame*

With these equations we deduce the relativistic transformation for the mean energy in a single *WEP* as measured by the two identical photo detectors each stationary in its reference frame using (9) and (10) as:

$$\frac{W'_{packet}}{W_{packet}} = \frac{T'}{T}\frac{E'^2_{max}}{E^2_{max}} = \frac{T'}{T}\cdot\frac{1-\beta}{1+\beta} = \frac{\lambda'}{\lambda}\cdot\frac{1-\beta}{1+\beta} \qquad (16)$$

Including also the equations for the relativistic Doppler shift deduced above $\frac{T'}{T} = \frac{\lambda'}{\lambda} = \sqrt{\frac{1+\beta}{1-\beta}}$ we have finally:

$$\frac{W'_{packet}}{W_{packet}} = \sqrt{\frac{1-\beta}{1+\beta}} \qquad (17)$$

which is the same equation as for the Lorentz transformation of energy. We further define a transversal energy density as the mean energy deposed by a *WEP* on the transversal unit area - $\rho_{s,packet} = \frac{W_{packet}}{A} = \lambda\varepsilon_0\frac{E^2_{max}}{4}$. As the transversal area is relativistic invariant we get further the relation:

$$\frac{\rho'_{s,packet}}{\rho_{s,packet}} = \sqrt{\frac{1-\beta}{1+\beta}} \qquad (18)$$

We conclude that in relativistic approach when seen by observers that move away from the source the *WEP* is subject to a couple of relativistic effects:
- A dilation of packet length (relativistic longitudinal dilation with the Doppler factor)
- A dilation of packet duration (relativistic time dilation with the Doppler factor)
- A transversal energy expansion, by which the energy deposed on the unit of transversal area decreases with the Doppler factor.

**4. The wave energy packet in the quantum theory approach**
In the spirit of quantum mechanics the wave packet is a mathematical solution for the Schrödinger equation. The square of the area under the wave packet solution is interpreted as the probability to locate some particle in a certain region. In quantum theory approach the wave energy is a discrete function of its frequency *E=N·h·v* or alternatively the accumulated energy of *N* photons each photon caring the same elementary energy quanta $E_{photon}=h\cdot v$. Based on this valuable knowledge and including the concept of the wave energy packet that we introduced previously we elaborate some new working hypothesizes:
- Because the energy density in plane wave oscillates periodically passing through zero we conclude that the wave energy travels as discrete photon groups. These groups correspond to the wave energy packet that we introduced earlier and they are consequently separated longitudinally by the period *λ/2* and temporary by the period *T/2*.

- The probability to detect a photon in the region of an wave energy packet approach zero at the packet limits (where the electric and magnetic field are both approaching zero) and increases to maximum in the middle of the packet (where the intensities of the electric and magnetic field are both reaching the local maximum).

In conclusion we may interpret the *WEP* as being the volumetric photon distribution function, i.e. the probability to find the photons carrying the wave energy in a certain region of space and at certain time. A photon counter placed somewhere along the propagation direction will record a counting rate oscillating periodically between a maximum that depends on the wave peak energy and a minimum that approaches zero. Consistently, a set of photon counters distributed along the propagation axis will record at a given time different instantaneous counting rates depending on their relative position. Photon counters separated by a multiple of half wavelength will record equal instantaneous counting rates that changes synchronously.

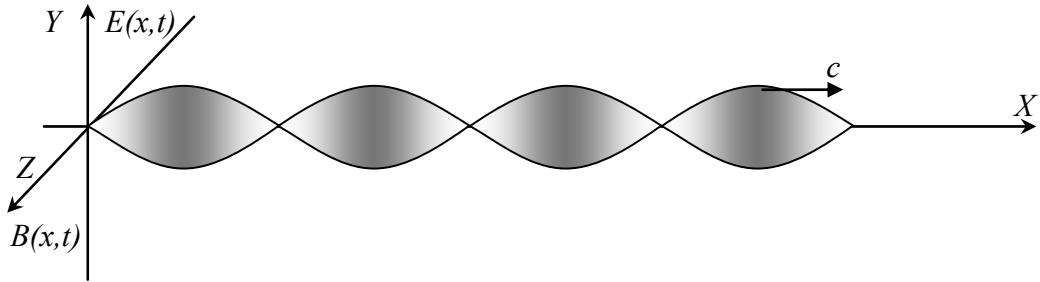

*Figure 5. Intuitive representation of the volumetric photon distribution by different gray intensities (darker gray represents higher photon density)*

With this hypothesis in mind we calculate the mean number of photons within a *WEP* as counted by the stationary photo detectors in figure 3. We have:

$$N = \frac{W_{packet}}{h\nu} \quad \text{and} \quad N' = \frac{W'_{packet}}{h\nu'} \tag{19}$$

These numbers are statistically significant if the experiment is repeated many times to get rid of the errors caused by quantum fluctuations. We find further that:

$$\frac{N'}{N} = \frac{W'_{packet}}{W_{packet}} \cdot \frac{T'}{T} = 1 \tag{20}$$

thus the average number of photons within an *WEP* is relativistic invariant. However, as we found in the prior section, the *WEP* mean energy decreases and its length increases. This means by a quantum approach that the photon probability distribution function stronger decreases - an effect of photon spreading. The conclusion is that even if the light speed is invariant, the photons may be sharply located by observers that are stationary in respect to the light source than the ones departing the source.

## 5. The wave energy packets and the wave volume

In equations (9-10) we introduce the notations $V_{packet} = \frac{\lambda}{2} A = \frac{cT}{2} A$ where $V_{packet}$ is the *WEP* volume corresponding to transversal area *A*. With this we have:

$$\frac{V'_{packet}}{V_{packet}} = \frac{\lambda'}{\lambda} = \sqrt{\frac{1+\beta}{1-\beta}} \qquad (21)$$

We conclude that the volume of unbounded wave energy packet is subject to Doppler dilation and not to Lorentz contraction as the volume of matter or material objects. Einstein himself was aware of the fact that length contraction applied to the transformation of the volume where the photons are located leads to the contradictory result that the counted number of photons is not a relativistic invariant and proposed a way out of it.[1,2,3]

## 8. The wave packet concept applied to the Margaritondo's experiment[4]

In this experiment a photo-detector stationary in a reference frame K counts the number of incident photons. The same experiment is contemplated by the observers of a second inertial reference frame K' that is moving at constant speed *V* relative to the first frame.

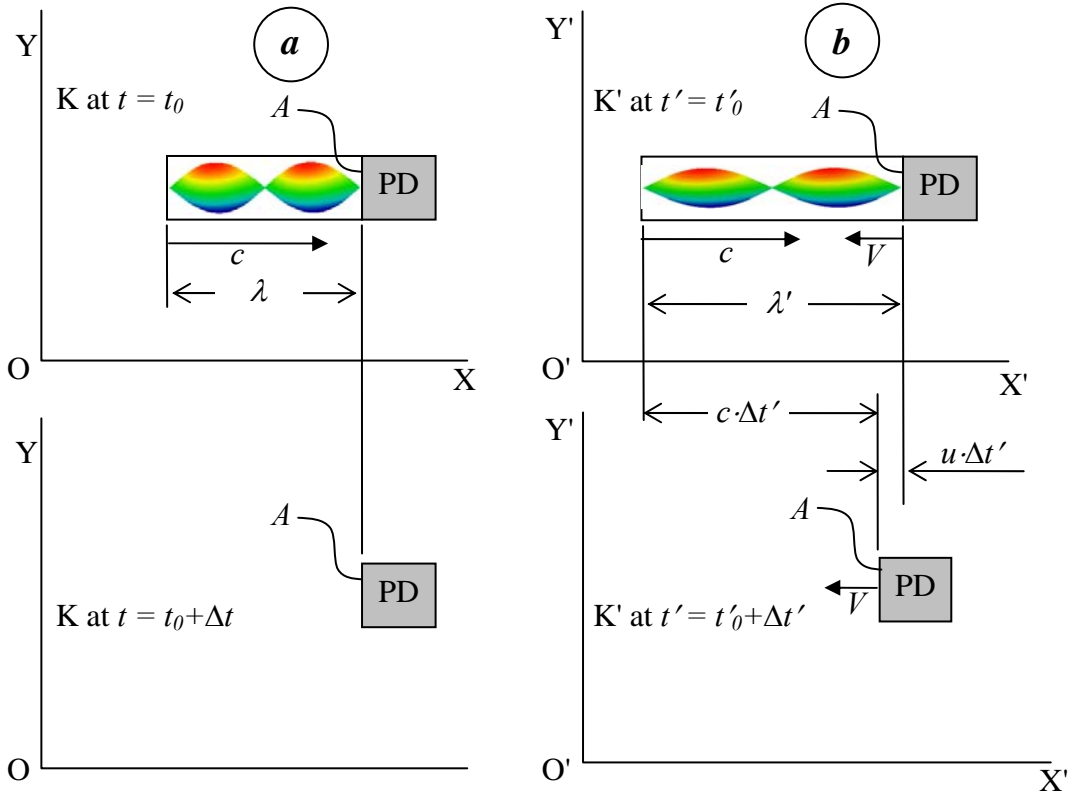

*Figure 4. Margaritondo's experiment used for deriving the transformation equation for the energy of a photon. Figure 4a depicts the experiment as detected from the rest frame of the photomultiplier whereas Figure 4b illustrates the experiment as detected from a reference frame that moves uniformly in respect to the first one.*

We accept that the number of photon counts is the same when viewed by the observers of the two inertial frames and we apply the concept of wave energy packet that we presented before. We know that the mean number of photons per wave packet is invariant. Therefore if the photo-detector is counting let say two complete wave packets in frame K then it shall count also two complete wave packets when seen from frame K' to get the same number of counts. We calculate the time window required to perform the counting process. In frame K we have $\Delta t = \frac{\lambda}{c}$ whereas in frame K' we have $\Delta t' = \frac{\lambda'}{V_{rel}}$ were $V_{rel}$ is the relative speed between the photo-detector and the wave packets as seen by observes in frame K'. This is the apparent speed between the wave packets (the wave front of the light signal) and a moving object in frame K' as seen from frame K'. It isn't the speed of light $c$ as properly seen by the photo-detector from its own reference frame. Using the Lorentz transformation for the measurement time intervals $\frac{\Delta t'}{\Delta t} = \frac{1}{\sqrt{1-\beta^2}}$ and the relativistic Doppler shift $\frac{\lambda'}{\lambda} = \sqrt{\frac{1+\beta}{1-\beta}}$ we obtain the final relation for the apparent relative speed as:

$$V_{rel} = c \cdot \frac{\lambda'}{\lambda} \cdot \frac{\Delta t}{\Delta t'} = c \cdot (1+\beta) = c + V \qquad (22)$$

which confirms the Margaritondo's assumption.

## 9. Conclusion

We introduced the concept of electromagnetic wave energy packet and analyzed the way this concept could be used to conciliate different physical approaches including the classical electromagnetic wave theory, the special relativity and the quantum theory. The length of unbounded wave packet doesn't transform as the length of the meter stick, which is subject to length contraction, but as the wavelength thus dilating with increasing velocity. This further sustains the proposal to measure the wave volume as discrete entities[5].